\documentclass[conference]{IEEEtran}
\usepackage{amssymb,color,amsmath}
\usepackage{paralist,cite}
\usepackage{graphicx}
\makeatletter
\def\ps@headings{%
\def\@oddhead{\mbox{}\scriptsize\rightmark \hfil \thepage}%
\def\@evenhead{\scriptsize\thepage \hfil \leftmark\mbox{}}%
\def\@oddfoot{}%
\def\@evenfoot{}}
\makeatother 

\newcommand{\nix}[1]{}
\newcommand{\F}{\mathbf{F}}
\newcommand{\Z}{\mathbf{Z}}
\newtheorem{theorem}{\textbf{Theorem}}
\newtheorem{lemma}[theorem]{\textbf{Lemma}}
\newtheorem{defn}[theorem]{\textbf{Definition}}

\newtheorem{example}[theorem]{\textbf{Example}}
\DeclareMathOperator{\ord}{ord}
\begin{document}

\title{Families of LDPC Codes Derived from Nonprimitive BCH Codes and Cyclotomic Cosets}
\author{
\authorblockN{Salah A. Aly}
\authorblockA{Department  of Computer Science \\
Texas A\&M University\\
College Station, TX 77843, USA \\
Email: salah@cs.tamu.edu}  }  \maketitle

\begin{abstract}
Low-density parity check (LDPC) codes are an important class of
codes with many applications. Two algebraic methods for constructing
regular LDPC codes are derived -- one based on nonprimitive
narrow-sense BCH codes and the other directly based on cyclotomic
cosets. The constructed codes have high rates and are free of cycles
of length four; consequently, they can be decoded using standard
iterative decoding algorithms. The exact dimension and bounds for
the minimum distance and stopping distance are derived. These
constructed codes can be used to derive quantum error-correcting
codes.
\end{abstract}

\section{Introduction}\label{sec:intro}
Bose-Chaudhuri-Hochquenghem (BCH) codes are  an interesting class of
linear codes that has been investigated for nearly  half of a
century. These types of codes have a rich algebraic structure.  BCH
codes with parameters $[n,k,d\geq \delta]_q$ are interesting because
one can choose their dimension $k$ and minimum distance $d$ once
given their design distance $\delta$ and length $n$ over a finite
field with $q$ elements. A linear code defined by a generator
polynomial $g(x)$ has dimension $k=n-deg(g(x))$ and rate $k/n$.  It
is not easy to show the dimension of nonprimitive BCH codes over
higher finite fields. In~\cite{aly06a,aly07a}, we have given an
explicit formula for the dimension of these codes if their deigned
distance $\delta$ is less than a constant $\delta_{\max}$.

Low-density parity check (LDPC) codes are a capacity-approaching
class of codes that were first described in a seminal work by
Gallager~\cite{gallager62}. Tanner in~\cite{tanner81} rediscovered
LDPC codes using a graphical interpretation.  A regular $(\rho,
\lambda)$ LDPC code is measured by the weights of its columns $\rho$
and rows $\lambda$. Iterative decoding algorithms of LDPC and turbo
codes highlighted the importance of these classes of codes for
communication and storage channels. Furthermore, these codes are
practical and have been used in many beneficial
applications~\cite{macKay98,lin04}. In contrast to BCH and
Reed-Solomon (RS) cyclic codes, LDPC cyclic codes with sparse parity
check matrices are customarily constructed by a computer search. In
practice, LDPC codes can achieve higher performance and better error
correction capabilities than many other codes, because they have
efficient iterative decoding algorithms, such as the product-sum
algorithm~\cite{tanner04,luby01,liva06,lin04}. Some BCH codes turned
out to be LDPC cyclic codes as well; for example, a $[15,7]$ BCH
code is also an LDPC code with a minimum distance five.

Regular and irregular LDPC codes have been constructed based on
algebraic and random approaches~\cite{song06,lin04}, and references
therein. Liva~\emph{et al.}~\cite{liva06} presented a survey of the
previous work done on algebraic constructions of LDPC codes based on
finite geometry, elements of finite fields, and RS codes.
Yi~\emph{et al.}~\cite{yi05} gave a construction for LDPC codes,
based on binary narrow-sense primitive BCH codes, and their method
is free of cycles of length four. Furthermore, a good construction
of LDPC codes should have a girth of the Tanner graph, of at least
$6$~\cite{liva06,lin04}. One might wonder how do the rates and
minimum distance of BCH codes compare to LDPC codes? Do
self-orthogonal BCH codes give raise to self-orthogonal LDPC codes
as well under the condition $\delta \leq \delta_{max}$. We  show
that how to derive LDPC codes from nonprimitive BCH codes.

One way to measure the decoding performance of linear codes is by
computing their~\emph{minimum distance} $d_{min}$. The performance
of low-density parity check codes under iterative decoding can also
be gauged by measuring their \emph{stopping sets} $S$
and~\emph{stopping distance} $s$, which is the size of the smallest
stopping set~\cite{schwartz06,orlitsky05}. For any given parity
check matrix $\textbf{H}$ of an LDPC code $\mathcal{C}$, one can
obtain the Tanner graph $G$ of this code and computes the stopping
sets. Hence, $s$ is a property of $\textbf{H}$, while $d_{min}$ is a
property of $\mathcal{C}$. The minimum distance is also bounded by
$d_{min} \geq s$. BCH codes are decoded invertible matrices such as
Berkcampe messay method, LDPC codes ar decoded using iterative
decoding  and Belief propagation (BP) algorithms.

In this paper,  we give a series of regular LDPC and Quasi-cyclic (QC)-LDPC
code constructions based on non-primitive narrow-sense BCH codes and elements
of cyclotomic cosets. The constructions are called \textbf{Type-I} and
\textbf{Type-II} regular LDPC codes. The algebraic structures of these codes
help us to predict additional properties of these codes. Hence, The constructed
codes have the following characteristics:
\begin{compactenum}[i)]
\item Two classes of regular  LDPC codes are constructed that have high rates and free of cycles of length
 four. Their properties can be analyzed easily.
\item The exact dimension is computed and the minimum distance is bounded for the constructed codes. Also, the
stopping sets and stopping distance can be determined from the
structure of the parity check matrices.
\end{compactenum}


The motivation for our work is to construct Algebraic regular LDPC
codes that can be used to derive quantum error-correcting codes.
Alternatively, they can also be used for wireless communication
channels. Someone will argue about the performance and usefulness of
the constructed regular LDPC codes in comparison to irregular LDPC
codes. Our first motivation is to derive quantum LDPC codes based on
nonprimitive BCH codes. Hence, the constructed LDPC-BCH codes can be
used to derive classes of symmetric quantum
codes~\cite{calderbank98,macKay04,aly07a,aly08c,aly08f} and
asymmetric quantum codes~\cite{evans07,steane96}. The literature
lacks many constructions of algebraic quantum LDPC codes, see for
example~\cite{macKay04,aly08c} and references therein.

\section{Constructing LDPC Codes}\label{sec:LDPC}
Let $\F_q$ denote a finite field of characteristic $p$ with $q$
elements. Recall that the set $\F_q^*=\F_q\setminus \{0\}$ of nonzero
field elements is a multiplicative cyclic group of order $q-1$.  A
generator of this cyclic group is called a primitive element of the
finite field $\F_q$.

\subsection{Definitions}
Let $n$ be a positive integer such that $\gcd(n,q)=1$ and $q^{\lfloor
m/2\rfloor} <n \leq \mu=q^m-1$, where $m=\ord_n(q)$ is the
multipicative order of $q$ modulo $n$.

Let $\alpha$ denote a fixed primitive element of~$\F_{q^m}$.  Define a map
$\textbf{z}$ from $\F_{q^m}^*$ to $\F_2^\mu$ such that all entries of
$\textbf{z}(\alpha^i)$ are equal to 0 except at position $i$, where it is equal
to 1. For example, $\textbf{z}(\alpha^2)=(0,1,0,\ldots,0)$.  We call
$\textbf{z}(\alpha^k)$ the location (or characteristic) vector of $\alpha^k$.
We can define the location vector $\textbf{z}(\alpha^{i+j+1})$ as the right
cyclic shift of the location vector $\textbf{z}(\alpha^{i+j})$, for $0 \leq j
\leq \mu-1$, and the power is taken module $\mu$.


\begin{defn}\label{def:Amatrix}We can define a map $A$ that associates to an element $\F_{q^m}^*$ a circulant
matrix in $\F_2^{\mu\times \mu}$ by
\begin{eqnarray}\label{label:mapA} A(\alpha^i)=\left ( \begin{array}{ccc}
\textbf{z}(\alpha^i) \\  \textbf{z}(\alpha^{i+1})
\\ \vdots \\  \textbf{z}(\alpha^{i+\mu-1})
\end{array} \right).
\end{eqnarray}
By construction, $A(\alpha^k)$ contains a 1 in every row and column.
\end{defn}

For instance,  $A(\alpha^1)$ is the identity matrix of size
$\mu \times \mu$, and $A(\alpha^2)$ is the shift matrix
\begin{eqnarray} A(\alpha^2)=\left ( \begin{array}{cccccc} 0&1&0&\ldots&0 \\
0&0&1&\ldots&0\\
\vdots&\vdots&\vdots&\vdots&\vdots\\1&0&0&\ldots&0
\end{array} \right).
\end{eqnarray}

We will use the map $A$ to associate to a parity check matrix $H=(h_{ij})$ in
$(\F_{q^m}^*)^{a\times b}$ the (larger and binary) parity check matrix
$\textbf{H}=(A(h_{ij}))$ in $\F_2^{\mu a\times \mu b}$. The matrices
$A(h_{ij})$$'s$ are $\mu \times \mu$ circulant permutation matrices based on
some primitive elements $h_{ij}$ as shown in Definition~\ref{def:Amatrix}.

\subsection{Regular LDPC Codes}
A low-density parity check code (or LPDC short) is a binary block code
that has a parity check matrix $\textbf{H}$ in which each row (and
each column) is sparse. An LDPC code is called \textit{regular} with
parameters $(\rho,\lambda)$ if it has a sparse parity check matrix $H$
in which each row has $\rho$ nonzero entries and each column has
$\lambda$ nonzero entries.

A regular LDPC code defined by a parity check matrix $\textbf{H}$ is
said to satisfy the \textit{row-column condition} if and only if any
two rows (or, equivalently, any two columns) of $\textbf{H}$ have at
most one position of a nonzero entry in common.  The row-column
condition ensures that the Tanner graph does not have cycles of
length four.

A Tanner graph of a binary code with a parity check matrix
$\textbf{H}=(h_{ij})$ is a graph with vertex set $V\stackrel{.}{\cup} C$ that
has one vertex in $V$ for each column of $\textbf{H}$ and one vertex in $C$ for
each row in $\textbf{H}$, and there is an edge between two vertices $i$ and $j$
if and only if $h_{ij}\neq 0$. Thus, the Tanner graph is a bipartite graph. The
vertices in $V$ are called the variable nodes, and the vertices in $C$ are
called the check nodes.  We refer to $d(v_i)$ and $d(c_j)$ as the degrees of
variable node $v_i$ and check node $c_j$ respectively.

Two values used to measure the performance of the decoding algorithms of LDPC
codes are: girth of a Tanner graph and stopping sets. The minimum stopping set
is analogous to the minimum Hamming distance of linear block codes.



\begin{defn}[Grith of a Tanner graph]\label{def:girth}
The girth $g$ of the Tanner graph is the length of its shortest cycle (minimum
cycle).
\end{defn}
A Tanner graph with large girth is desirable, as iterative decoding converges
faster for graphs with large girth.


\begin{defn}[Stopping set]\label{defn:stopset}
A \textit{stopping set} $S$ of a Tanner graph is a subset of the variable nodes
$V$ such that each vertex in the neighbors of $S$ is connected to at least two
nodes in $S$.
\end{defn}

The \textit{stopping distance} is the size of the smallest stopping set. The
stopping distance determines the number of correctable erasures by an iterative
decoding algorithm, see~\cite{orlitsky05,schwartz06,di00}.

\begin{defn}[Stopping distance]
The stopping distance of the parity check matrix $\textbf{H}$ can be defined as
the largest integer $s(\textbf{H})$ such that every set of at most
$(s(\textbf{H})-1)$ columns of \textbf{H} contains at least one row of weight
one, see~\cite{schwartz06}.
\end{defn}
 The stopping ratio $\sigma$ of the Tanner graph of
a code of length $n$ is defined by $s$ over the code length.

The minimum Hamming distance is a property of the
code used to measure its performance for maximum-likelihood
decoding, while the stopping distance is a property of the parity
check matrix $\textbf{H}$ or the Tanner graph $G$ of a specific
code. Hence, it varies for different choices of $\textbf{H}$ for the
same code $\mathcal{C}$.  The stopping distance $s(\textbf{H})$ gives
a lower bound of the minimum distance of the code $\mathcal{C}$
defined by \textbf{H}, namely
\begin{eqnarray}
 s(\textbf{H}) \leq d_{min}
\end{eqnarray}
It has been shown that finding the stopping sets of minimum
cardinality is an NP-hard problem, since the minimum-set vertex
covering problem can be reduced to it~\cite{krishnan07}.


\section{LDPC Codes based on  BCH Codes}\label{sec:LDPC-BCH}
In this section we give two constructions of LDPC codes derived from
nonprimitive BCH codes, and from elements of cyclotomic cosets.
In~\cite{yi05}, the authors derived a class of regular LDPC codes
from primitive BCH codes but they did not prove that the
construction has free of cycles of length four in the Tanner graph.
In fact, we will show that not all primitive BCH codes can be used
to construct LDPC with cycles greater than or equal to six in their
Tanner graphs. Our construction is free of cycles of length four if
the BCH codes are chosen with prime lengthes as proved in
Lemma~\ref{lemma:freecycles4}; in addition the stopping distance is
computed. Furthermore, We are able to derive a formula for the
dimension of the constructed LDPC codes as given in
Theorem~\ref{thm:Hrank}. We also infer the dimension and cyclotomic
coset structure of the BCH codes based on our previous results
in~\cite{aly06a,aly07a}.

We keep the definitions of the previous section. Let $q$ be a power of
a prime and $n$ a positive integer such that $\gcd(q,n)=1$. Recall
that the cyclotomic coset $C_x$ modulo $n$ is defined as
\begin{eqnarray}C_x=\{xq^i\bmod n \mid i\in \Z, i\ge 0\}.
\end{eqnarray}

Let $m$ be the multiplicative order of $q$ modulo $n$. Let $\alpha$ be
a primitive element in $\F_{q^m}$. A nonprimitive narrow-sense BCH
code $\mathcal{C}$ of designed distance $\delta$ and length $n$ over
$\F_{q}$ is a cyclic code with a generator monic polynomial $g(x)$
that has $\alpha, \alpha^2, \ldots, \alpha^{\delta-1}$ as zeros,
\begin{eqnarray}
g(x)=\prod_{i=1}^{\delta -1} (x-\alpha^i).
\end{eqnarray}
Thus,  $c$ is a codeword in $\mathcal{C}$ if and only if
$c(\alpha)=c(\alpha^2)=\ldots=c(\alpha^{\delta-1})=0$. The parity check matrix
of this code can be defined as
\begin{eqnarray}\label{bch:parity}
 H_{bch} =\left[ \begin{array}{ccccc}
1 &\alpha &\alpha^2 &\cdots &\alpha^{n-1}\\
1 &\alpha^2 &\alpha^4 &\cdots &\alpha^{2(n-1)}\\
\vdots& \vdots &\vdots &\ddots &\vdots\\
1 &\alpha^{\delta-1} &\alpha^{2(\delta-1)} &\cdots &\alpha^{(\delta-1)(n-1)}
\end{array}\right].
\end{eqnarray}

We note the following fact about the cardinality of cyclotomic cosets.

\begin{lemma}\label{th:bchnpcosetsize}
Let $n$ be a positive integer and $q$ be a power of a prime, such that
$\gcd(n,q)=1$ and $q^{\lfloor m/2\rfloor} <n \leq q^m-1$, where $m=ord_n(q)$.
The cyclotomic coset $C_x=\{ xq^j\bmod n \mid 0\le j<m\}$ has a cardinality of
$m$ for all $x$ in the range $1\leq x\leq nq^{\lceil m/2\rceil}/(q^m-1).$
\end{lemma}
\begin{proof}
See~\cite[Lemma 8]{aly07a}.
\end{proof}

Therefore, all cyclotomic cosets have the same size $m$ if their
range is bounded by a certain value. This lemma enables one to
determine the dimension in closed form for BCH code of small
designed distance~\cite{aly06a,aly07a}. In fact, we show the
dimension of nonprimitve BCH codes over $\F_q$.
%

\begin{theorem}\label{th:bchnpdimension}
Let $q$ be a prime power and $\gcd(n,q)=1$, with $ord_n(q)=m$. Then a
narrow-sense BCH code of length $q^{\lfloor m/2\rfloor} <n \leq q^m-1$ over
$\F_q$ with designed distance $\delta$ in the range $2 \leq \delta \le
\delta_{\max}= \min\{ \lfloor nq^{\lceil m/2 \rceil}/(q^m-1)\rfloor,n\}$, has
dimension of
\begin{equation}\label{eq:npdimension}
k=n-m\lceil (\delta-1)(1-1/q)\rceil.
\end{equation}
\end{theorem}
\begin{proof}
See~\cite[Theorem 10]{aly07a}.
\end{proof}

Based on these two observations, we can construct regular LDPC codes from BCH
codes with a known dimension and cyclotomic coset size.

\subsection{ {Type-I Construction}}
In this construction, we use the parity check matrix of a nonprimitive
narrow-sense BCH code over $\F_q$ to define the parity check
matrix of a regular LDPC over $\F_2$.

Consider the narrow-sense BCH code of prime length $q^{\lfloor
m/2\rfloor} <n \leq q^m-1$ over $\F_q$ with designed distance
$\delta$ and  $ord_n(q)=m$. We use the fact that there must be some
primes in the integer range $(q^{\lfloor m/2\rfloor}, q^m-1)$. In
fact, there must exist a prime between $x$ and $2x$ for some integer
x, in which it ensures existence primes in the given interval. A
parity check matrix $\textbf{H}$ of an LDPC code can be obtained by
applying the map $A$ in Equation~(\ref{label:mapA}) to each entry of
the parity check matrix~(\ref{bch:parity}) of this BCH code,
\begin{eqnarray}\label{eq:LDPCtype-I} &\textbf{H}&=\\ &&\!
\left[ \begin{array}{ccccc}
A(1) &A(\alpha)&A(\alpha^2)\!\!\!&\cdots\!\!\!&\!\!\! A(\alpha^{n-1})\!\!\!\\
A(1) &A(\alpha^2)&A(\alpha^4 )\!\!\!&\cdots\!\!\!&\!\!\! A(\alpha^{2(n-1)})\!\!\!\\
\vdots& \vdots &\vdots &\ddots\!\!\!  &\!\!\! \vdots\!\!\! \\
A(1) &A(\alpha^{\delta-1})\!\!\!&A(\alpha^{2(\delta-1)}
)&\cdots&\!\!\!A(\alpha^{(\delta-1)(n-1)}) \!\!\!
\end{array}\right].\nonumber
\end{eqnarray}
The matrix $\textbf{H}$ is of size $(\delta-1)\mu \times n\mu$ and by
construction it has the following properties:
\begin{compactitem}
\item Every column has a weight of  $\delta-1$.
\item Every row has a weight of $n$.
\end{compactitem}

The matrix $\textbf{H}$ of size $(\delta-1)\mu \times n\mu$ has a
weight of $\rho=\delta-1$ in every column, and a weight of
$\lambda=n$ in every row. The null space of the matrix $\textbf{H}$
defines a $(\rho,\lambda)$ LDPC code with a high rate for a small
designed distance $\delta$ as we will show. The minimum distance of
the BCH code is bounded by
\begin{eqnarray}
d_{min} \geq \left\{
               \begin{array}{ll}
                 \delta+1, & \hbox{odd $\delta$;} \\
                 \delta+2, & \hbox{even $\delta$.}
               \end{array}
             \right.
\end{eqnarray}
Also, the minimum distance of the LDPC codes is bounded by
$d_{min}$.
Now, we will show that in general  regular $(\rho,\lambda)$ LDPC
codes derived from  primitive BCH codes of length $n$  are not free
of cycles of length four as claimed in~\cite{yi05}.
%
%


\begin{lemma}\label{lemma:freecycles4}
The Tanner graph of LDPC codes constructed in~\textbf{Type-I} are
free of cycles of length four for a prime length $n$.
\end{lemma}
\begin{proof}
Consider the block-column indexed by $n-j$ for $1 \leq j \leq n-1$
and let $r_i$ and $r_i'$ be two different block-rows  for $1 \leq
r_i,r_i' \leq (\delta-1)$. Assume by contradiction that we have
$A(\alpha^{r_i(n-j)}) = A(\alpha^{r_j'(n-j)})$. Thus $r_i(n-j) \mod
n=r_i'(n-j) \mod n$  or $n(r_i-r_i') \mod n=(r_i-r_i')j \mod n=0$.
This contradicts the assumption that $n> j \geq 1$ and $r_i\neq
r_i'$.
\end{proof}
Hence primitive BCH codes of composite length $n$ can not be used to
derive LDPC codes that are cycles-free of length four using our
construction.

%
The proof of the following lemma is straight forward by  exchanging,
adding, and permuting a  block-row.

\begin{lemma}\label{lem:twoblocks}
Let $(\ldots,1_\ell,\ldots)$ be a vector of length $\mu$ that has 1 at position
$\ell$. Under the cyclic shift, the following two blocks $h_a$ and $h_b$ of
size $\mu \times \mu$  are equivalent, where $h_a$ and $h_b$ are generated by
the rows $\left(\begin{array}{ccccc} 1&\ldots&1_i&\ldots\end{array}\right)$ and
$\left(\begin{array}{ccccc} 1&\ldots&1_j&\ldots\end{array}\right)$ and their
cyclic shifts, respectively.
\end{lemma}

One might imagine that the rank of the parity check matrix
$\textbf{H}$ in~(\ref{eq:LDPCtype-I}) is given by $(\delta-1)\mu$
since rows of every block-row $h_a$ is linearly independent. A
computer program has been written to check the exact formula and
then we drove a formula to give the rank of the matrix $\textbf{H}$.
%

\begin{theorem}\label{thm:Hrank}
Let $n$ be a prime in the range $q^{\lfloor m/2\rfloor} <n \leq
\mu=q^m-1$ and $\delta$ be an integer in the range $2 \leq \delta <
n$ for some prime power $q$ and $m=\ord_q(n)$. The rank of the
parity check matrix $\textbf{H}$ given by

\begin{eqnarray}\label{eq:LDPCtype-I} \textbf{H}=\!
\left[ \begin{array}{ccccc}
\mathcal{A}^o &\mathcal{A}^1&\mathcal{A}^2\!\!\!&\cdots\!\!\!&\!\!\! \mathcal{A}^{n-1}\!\!\!\\
\mathcal{A}^0 &\mathcal{A}^2&A^4\!\!\!&\cdots\!\!\!&\!\!\! \mathcal{A}^{2(n-2)}\!\!\!\\
\vdots& \vdots &\vdots &\ddots\!\!\!  &\!\!\! \vdots\!\!\! \\
\mathcal{A}^0 &\mathcal{A}^{\delta-1}\!\!\!&\mathcal{A}^{\delta-1}
)&\cdots&\!\!\!\mathcal{A}^{(\delta-1)(n-1)} \!\!\!
\end{array}\right]
\end{eqnarray}
 is
$(\delta-1)\mu-(\delta-2)$, where $\mathcal{A}^i=A(\alpha^i )$.
\end{theorem}
\begin{proof}
The proof of this theorem can be shown by mathematical induction for
$1,2,\ldots,\delta \leq n$. We know that every block-row is linearly
independent.
\begin{compactenum}[i)]
\item Case i. Let $\delta=2$, the statement is true since ever block-row has
only 1 in every column, the first n columns represent the identity matrix.

\item Case ii-1. Assume the statement is true for $\delta-2$. In this case,  the matrix
\textbf{G} has a full rank given by $(\delta-2)\mu-(\delta-3)$.  So, we have
$$\textbf{G}=\left(\begin{array}{cccccc} h_{11}&h_{12}&h_{13}&\ldots&\ldots&h_{1n}\\0&h_{22}&h_{23}&\ldots&\ldots&h_{2n} \\
0&0&h_{33}&\ldots&\ldots&h_{3n}\\ 0&0&0&\vdots&\vdots&h_{in}\\
0&0&\ldots&h_{(\delta-2) (\delta-2)}&\ldots&h_{(\delta-2)n}
\end{array}\right).$$
The elements $h_{ii}'s$ have 1's in the diagonal and zeros
everywhere using simple Gauss elimination method and
Lemma~\ref{lem:twoblocks}.

\item Case iii-1.
We can form the sub-matrix $\textbf{H}_2$ of size $(\delta-1)\mu
\times (\delta-1)\mu$ by adding one block-row to the matrix
$\textbf{G}$. The last block-row is generated by
$$(A(\alpha^0),A(\alpha^{\delta-1}),A(\alpha^{2(\delta-1)}),\ldots,A(\alpha^{n-1(\delta-1)})).$$
 All $\mu-1$ rows of the last block-row are linearly independent
and can not be generated from the previous $\delta-2$ blocks-row.
Now, in order to obtain the last row-block to be zero at positions
$h_{(\delta-1)1},h_{(\delta-1)2},\ldots,h_{(\delta-1)(\delta-2)}$,
we can add the element $h_{jj}$ to the element $h_{(\delta-1)j}$. In
addition, the last row (row indexed by $(\delta-1)\mu$) of block-row
$\delta-1$ can be generated by adding all elements of the first
block-row  to the first $\mu-1$ rows of the last block-row.
$$\textbf{G}=\left(\begin{array}{cccccc} h_{11}&h_{12}&h_{13}&\ldots&\ldots&h_{1n}\\0&h_{22}&h_{23}&\ldots&\ldots&h_{2n} \\
0&0&h_{33}&\ldots&\ldots&h_{3n}\\ 0&0&0&\vdots&\vdots&h_{in}\\
0&0&\ldots&h_{(\delta-1) (\delta-1)}&\ldots&h_{(\delta-1)n}
\end{array}\right).$$

 Therefore, the matrix $\textbf{G}$ has rank of
$(\delta-2)\mu-(\delta-3)+\mu-1=(\delta-1)\mu-(\delta-2)$. We notice that the
matrix $\textbf{H}$ has the same rank as the matrix $\textbf{G}$, hence the
proof is completed.

\end{compactenum}
\end{proof}
The proof can also be shown by dropping the last row of every
block-row except at the last row in the first block-row. Hence, the
remaining matrix has a full rank.
Obtaining a formula for  rank of the parity check matrix \textbf{H}
allows us to compute rate of the constructed LDPC codes.  Now, we
can deduce the relationship between nonprimitive narrow-sense BCH
codes and LDPC codes constructed in~\textbf{Type-I}.


\begin{theorem}[LDPC-BCH Theorem]\label{lem:BCHtype-I}
Let $n$ be a prime and $q$ be a power of a prime, such that
$\gcd(n,q)=1$ and $q^{\lfloor m/2\rfloor} <n \leq q^m-1$, where
$m=ord_n(q)$. A nonprimitive narrow-sense BCH code with parameters
$[n,k,d_{min}]_q$ gives  a $(\delta-1,n)$ LDPC code with rate
$(n\mu-[(\delta-1)\mu-(\delta-2)])/n\mu$, where $k=n-m\lceil
(\delta-1)(1-1/q)\rceil$ and $2 \leq \delta \leq \delta_{max}$. The
constructed codes are free of cycles with length four.
\end{theorem}
\begin{proof}
By \textbf{Type-I} construction of LDPC codes derived from
nonprimitive BCH codes using Equation~(\ref{eq:LDPCtype-I}), we know
that every element $\alpha^i$ in $H_{bch}$ is a circulant matrix
$A(\alpha^i)$ in $\textbf{H}$. Therefore, there is a parity check
matrix $\textbf{H}$ with size $(\delta-1)\mu \times n\mu$.
$\textbf{H}$ has a row weight of $n$ and a column weight of
$\delta-1$. Hence, the null space of the matrix $\textbf{H}$ defines
an LDPC code with the given rate using Lemma~\ref{thm:Hrank}.

The constructed code is free of cycles of length four, because the
matrix $H_{bch}$ has no two rows with the same value in the same
column, except in the first column. Hence, the matrix $\textbf{H}$
has, at most, one position in common between two rows due to
circulant property and Lemma~\ref{lemma:freecycles4}. Consequently,
they have a Tanner graph with girth greater than or equal to six.
\end{proof}


Based on \textbf{Type-I} construction of regular LDPC codes, we
notice that every variable node has a degree $\delta-1$ and every
check nodes has a degree $n$. Also, the maximum number of columns
that do not have one in common is $n$. Therefore, the following
Lemma counts the stopping distance of the Tanner graph defined by
$\textbf{H}$.


\begin{lemma}\label{lem:stoppingset}
The cardinality of the smallest stopping set of the Tanner graph of
\textbf{Type-I} construction of regular LDPC codes is $\mu+1$.
\end{lemma}
%


\begin{example}
Let $n=\mu=q^m-1$, with $m=7$ and $q=2$.  Consider a BCH code with $\delta=5$
and length $n$. Assume $\alpha$ to be a primitive element in $\F_{q^m}$.  The
matrix $H$ can be  written as
\begin{eqnarray} H=\left ( \begin{array}{ccccccc} 1&\alpha&\alpha^2& \ldots&\alpha^{126} \\
1&\alpha^2&\alpha^4&\ldots&\alpha^{125}\\
1&\alpha^3&\alpha^6&\ldots&\alpha^{124}\\
1&\alpha^4&\alpha^8&\ldots&\alpha^{123}\\
\end{array} \right),
\end{eqnarray}
and the matrix $\textbf{H}$ has size $ 508 \times 16129 $. Therefore, we
constructed a $(4,127)$ regular LDPC with a rate of $123/127$, see
Fig.~\ref{fig:ldpc1}.
\end{example}

\begin{table}[ht]
\caption{Parameters of LDPC codes derived from NP BCH codes}
\label{table:bchtable}
\begin{center}
\begin{tabular}{|l|l|c|c|c|}
\hline   $q$ &$\mu$&  BCH Codes & LDPC code&rank of \textbf{H}\\&&&size of \textbf{H}    &\\
 \hline
 2&31&$[23,12,4]$&(93,713)&91\\
 3&26&$[23,12,5]$&(104,598)& 101 \\
 2&31&$[31,26,3]$&(62,961)&61\\
 2&31&$[31,21,5]$&(124,961,)&121\\
 2&31&$[31,26,6]$&(155, 961)&151\\
  2&31&$[31,16,7]$&(186,961)&181\\
    2&63&$[47,24,4]$&(189 ,1961)&187\\
  2&63&$[61,21,6]$&(315, 3843)& 311\\
  2 &63&$[61,11,10]$&(567,3843)&559\\
    2 &127&$[127,113,15]$&(1778,16129)&1765\\
2 &127&$[127,103,25]$&(3048,16129)&3025\\
&&&&\\
 \hline
\end{tabular}
\end{center}
\end{table}
%
\section{LDPC Codes Based on Cyclotomic Cosets} In this section we will construct regular
LDPC codes based on the structure of cyclotomic cosets. Assume that
we use the same notation as shown in Section~\ref{sec:LDPC}. Let
$C_x$ be a cyclotomic coset modulo prime integer $n$, defined as
$C_x=\{xq^i \bmod n \mid i\in \Z, 1 \leq x < n \}.$ We can also
define the location vector $\textbf{y}$ of a cyclotomic coset $C_x$,
instead of the location vector $\textbf{z}$ of an element
$\alpha^i$.
%


\begin{defn}
The location vector $\textbf{y}(C_x)$ defined over a cyclotomic coset $C_x$ is
the vector $\textbf{y}(C_x)=(z_0,z_1,\dots,z_n)$, where all positions  are
zeros except at positions corresponding to elements of $C_x$.
\end{defn}
Let $\ell$ be the number of different cyclotomic cosets $C_{x}^i$'s that are
used to construct the matrices $H_{C_j}^i$'s. We can index the $\ell$ location
vectors   corresponding to $C_{x_1},C_{x_2},\ldots,C_{x_\ell}$, as
$\textbf{y}^1,\textbf{y}^2,\ldots,\textbf{y}^\ell$.  Let $\textbf{y}^1(\gamma
C_x)$ be the cyclic shift of  $\textbf{y}^1(C_x) $ where every element in $C_x$
is incremented by 1.

\subsection{Type-II Construction}
 We construct the
matrix $H_{C_x}^1$ from the cyclotomic $C_x$ as
\begin{eqnarray}\label{eq:HcycloBCH}
 H_{C_x}^1=\left ( \begin{array}{ccc} \textbf{y}^1(C_x) \\  \textbf{y}^1(\gamma C_x)
\\ \vdots \\   \textbf{y}^1(\gamma^{n-1} C_x) \end{array} \right), \end{eqnarray}
where $  \textbf{y}^1(\gamma^{j+1} C_x)$ is the cyclic shift of  $
\textbf{y}^1(\gamma^{j} C_x) $ for $0\leq j \leq n-1$.

 From Lemma~\ref{th:bchnpcosetsize}, we know that all cyclotomic cosets $C_x$'s
 have a size of $m$ if $1\leq x\leq nq^{\lceil m/2\rceil}/(q^m-1).$

We can generate all rows of $H_{C_x}$, by shifting the first row one position
to the right. Our construction of the matrix $H_{c_x}^i$ has the following
restrictions.
\begin{compactitem}
\item Let $x \leq \Theta(\sqrt{n})$, this will guarantee that all
cyclotomic cosets have the same size $m$.
\item Any two rows of  $H_{c_x}^i$ have only one nonzero position in common.
\item Every row (column) in $H_{c_x}^i$ has a weight of $m$.
\end{compactitem}

We can construct the matrix $\textbf{H}$ from different cyclotomic cosets as
follows.
\begin{small}
\begin{eqnarray}
\textbf{H} &=& \Big[ \begin{array}{cccc}H_{C_1}^1&
H_{C_3}^2&\ldots&H_{C_j}^\ell
\end{array}\Big]\\&=&\left ( \begin{array}{ccccc}
\textbf{y}^1(C_1)&  \textbf{y}^2(C_2)&\ldots& \textbf{z}^\ell(C_j)\\
  \textbf{y}^1( \gamma C_1)&\textbf{y}^2(\gamma  C_2)&\ldots&
\textbf{y}^\ell(\gamma C_j)
\\ \vdots &\vdots&\vdots&\vdots \\\textbf{y}^1(\gamma^{n-1}  C_1)& \textbf{y}^2(\gamma^{n-1} C_2)&\ldots& \textbf{y}^\ell(\gamma^{n-1} C_j) \end{array} \right), \nonumber\end{eqnarray}
\end{small}
where we choose  the number $\ell$ of different sub-matrices $H_{C_j}$. The
$n\times (\ell*n)$ matrix $\textbf{H}$ constructed in \textbf{Type-II} has the
following properties.
\begin{compactenum}[i)]

\item Every column has a weight of $m$ and every row has a weight of $m*\ell$, where $\ell$ is
the  number of matrices $H_{C_j}'s$.
\item For a large n, the matrix $\textbf{H}$ is a sparse low-density parity check matrix.
\end{compactenum}
 We can also show that the null space  of the matrix $\textbf{H}$
defines an $(m,m\ell)$ LDPC code with rate $(\ell-1)/\ell$. Clearly,
an increase in $\ell$, increases the rate of the code.
%

Since all cyclotomic cosets $C_{x_1},C_{x_2},\ldots,C_{x_\ell}$ used to
construct \textbf{H} are different, then  the first column in each sub-matrix
$H_{C_x}^j$ is different from the first column in all sub-matrices  $H_{C_x}^i$
for $j\neq i$ and $1 \leq i \leq \ell$. Now, we can give a lower bound in the
stopping distance of \textbf{Type-II} LDPC codes.


\begin{lemma}
The stopping distance of LDPC codes, that are in \textbf{Type-II} construction,
is at least $\ell+1$.
\end{lemma}

 One can improve this bound, by counting the number of columns in each
sub-matrix $H_{C_x}^i$ that do not have one in common in addition to all
columns in the other sub-matrices.


\begin{example}
Consider $n=q^m-1$ with $m=5$, $q=2$, and $\delta=5$.  We can compute the
cyclotomic cosets $C_1$, $C_3$ and $C_5$ as $C_1=\{1,2,4,8,16\},$
$C_3=\{3,6,12,24,17\}$ and $C_5=\{5,10,20,9,18 \}$. The matrices $H_{C_1}^1$,
$H_{C_3}^2$ and $H_{C_5}^3$ can be defined based on $C_1$, $C_3$ and $C_5$,
respectively.
\begin{small}\begin{eqnarray} H_{C_1}^1=\left (
\begin{array}{ccccccccccccccccccccccccccccccccccccccccccc}
\!\!\!  1101 & \!\!\!0001 &\!\!\! 0000 &\!\!\! 0001 &\!\!\! 0000 &\!\!\! 0000 &\!\!\! 0000 &\!\!\! 000 \!\!\! \\
\!\!\!  0110 & \!\!\!1000 &\!\!\! 1000 &\!\!\! 0000 &\!\!\! 1000 &\!\!\! 0000 &\!\!\! 0000 &\!\!\! 000 \!\!\! \\
\!\!\!  0011 & \!\!\!0100 &\!\!\! 0100 &\!\!\! 0000 &\!\!\! 0100 &\!\!\! 0000 &\!\!\! 0000 &\!\!\! 000 \!\!\! \\
\!\!\!  0001 & \!\!\!1010 &\!\!\! 0010 &\!\!\! 0000 &\!\!\! 0010 &\!\!\! 0000 &\!\!\! 0000 &\!\!\! 000 \!\!\! \\
\!\!\!  0000 & \!\!\!1101 &\!\!\! 0001 &\!\!\! 0000 &\!\!\! 0001 &\!\!\! 0000 &\!\!\! 0000 &\!\!\! 000 \!\!\! \\
\!\!\!  \vdots & \!\!\! \vdots &\!\!\! \vdots &\!\!\! \vdots &\!\!\! \vdots &\!\!\! \vdots &\!\!\! \vdots &\!\!\! \vdots \!\!\! \\
\!\!\!  0100 & \!\!\!0100 &\!\!\! 0000 &\!\!\! 0100 &\!\!\! 0000 &\!\!\! 0000 &\!\!\! 0000 &\!\!\! 011 \!\!\! \\
\!\!\!  1010 & \!\!\!0010 &\!\!\! 0000 &\!\!\! 0010 &\!\!\! 0000 &\!\!\! 0000 &\!\!\! 0000 &\!\!\! 001 \!\!\! \\
\end{array} \right)
\end{eqnarray}
\end{small}

The matrix $\textbf{H}$ of size (31,93) is given by \begin{eqnarray} \textbf{H}
= \Big[
\begin{array}{cccc}H_{C_1}^1& H_{C_3}^2&H_{C_5}^3
\end{array}\Big],
\end{eqnarray} therefore, the null space of $\textbf{H}$ defines an (5,15) LDPC
code with parameters $(62,93)$.
\end{example}


We note that \textbf{Type-I} and \textbf{Type-II} constructions can
be used to derive quantum codes, if the parity check matrix
\textbf{H} is modified to be self-orthogonal or using the nested
propery of LDPC-BCH codes. Recall that quantum error-correcting
codes over $\F_q$ can be constructed from self-orthogonal classical
codes over $\F_q$ and $\F_{q^2}$, see for
example~\cite{aly07a,calderbank98,hagiwara07,macKay04,aly08f} and
references therein. \nix{In our future research, we plan to derive
quantum LDPC codes from \textbf{Type-I} and \textbf{Type-II}
constructions that are based on nonprimitve BCH codes.}

\section{Simulation Results}\label{sec:simulation}

\begin{figure}[t]
  \includegraphics[scale=0.45]{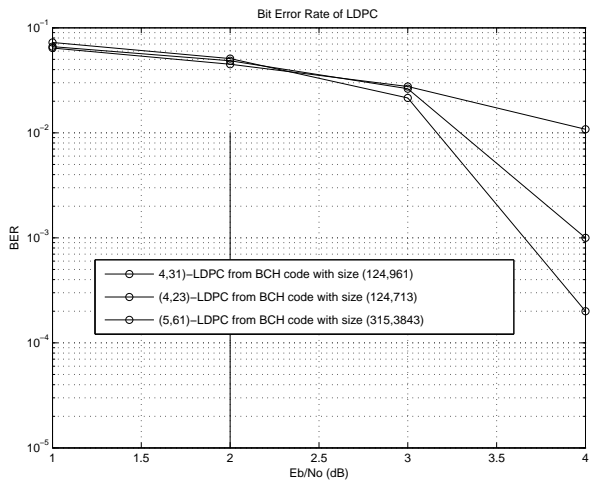}
 \centering
  \caption{\textbf{Type I:} The error performance of an (4,31) LDPC code with rate $27/31$ and H matrix with size $(124,961)$ based on a BCH code.}\label{fig:ldpc1}
\end{figure}

We simulated the performance of the constructed codes using standard
iterative decoding algorithms. Fig.~\ref{fig:ldpc1} shows the BER
curve for an (4,31) LDPC code~\textbf{Type I} with a length of 961,
dimension of 837, and  number of iterations  of 50. This performance
can also be improved for various lengths and the designed distance
of BCH codes.  The performance of these constructed codes can be
improved for large code length in comparison to other LDPC codes
constructed in~\cite{lin04,liva06}. As shown in Fig.~\ref{fig:ldpc1}
at the $10^{-4}$ BER, the code performs at $5.5$ $Eb/No(dB)$, which
is $1.7$ units from the Shannon limit.


\section{Conclusion}\label{sec:conclusion}
We introduced  two families of regular LDPC codes based on
nonprimitive narrow-sense BCH codes and structures of cyclotomic
cosets. We gave a systematic method to write every element in the
parity check matrix of BCH codes as vector of length $\mu$.  We
demonstrated that these constructed codes have high rates and a
uniform structure that made it easy to compute their dimensions,
stopping distance, and bound their minimum distance. Furthermore,
one can use standard iterative decoding algorithms to decode these
codes.\nix{ we plan to investigate more properties of these codes
and evaluate their performance over different communication
channels.} One can easily derive irregular LDPC codes based on these
codes and possibly increase performance of the iterative decoding.
Also, in future research, these constructed codes can be used to
derive quantum LDPC error-correcting codes.

\section*{Acknowledgments.}
Part of this work was accomplished during a research visit at
Bell-Labs \& Alcatel-Lucent in  Summer 2007. I thank my teachers,
colleagues, and family.

\emph{"Accurate reckoning: The entrance into knowledge of all
existing things and all obscure secrets." Foundation of true
science, Ahmes, Anc. EG. Scribe, 2000 BC. S.A.A. confirms that the
simple work accomplished in this paper is based on accurate
counting.}


\bibliographystyle{ieeetr}

\end{document}